\documentstyle[aps,prb]{revtex}

\begin{document}
\draft

\twocolumn[\hsize\textwidth\columnwidth\hsize\csname@twocolumnfalse\endcsname

\title{Decay properties of the one-particle Green function in real space and
imaginary time}
\author{Arno Schindlmayr\cite{byline}}
\address{Fritz-Haber-Institut der Max-Planck-Gesellschaft, Faradayweg 4--6,
14195 Berlin-Dahlem, Germany}
\date{\today}
\maketitle

\begin{abstract}
The decay properties of the one-particle Green function in real space and
imaginary time are systematically studied for solids. I present an analytic
solution for the homogeneous electron gas at finite and at zero temperature as
well as asymptotic formulas for real metals and insulators that allow an
analytic treatment in electronic-structure calculations based on a space-time
representation. The generic dependence of the decay constants on known
system parameters is used to compare the scaling of reciprocal-space
algorithms for the GW approximation and the space-time method.
{\em Copyright 2000 by The American Physical Society.}
\end{abstract}

]

\section{Introduction}

The GW approximation\cite{Hedin1965} for the electronic self-energy is
known to produce quasiparticle band structures in very good agreement with
experimental data for a wide range of crystalline materials that include
semiconductors,\cite{Hybertsen1985,Godby1986} simple
metals,\cite{Northrup1987} and even transition metals.\cite{Aryasetiawan1992}
It corresponds to a summation of ring diagrams to infinite order and may be
thought of as an extended Hartree-Fock scheme with dynamically screened
exchange. Despite the long-recognized success for bulk systems there are
still comparably few applications to more complicated geometries, however,
because of the rapid increase in computational cost. Unless additional
simplifications like a plasmon-pole model\cite{Hybertsen1985} for the screened
Coulomb interaction are employed, conventional reciprocal-space algorithms
scale like $N^4/b^2$ for a calculation of the complete spectral function,
where $N$ is a measure of the system size, such as the number of atoms in a
bulk material, and $b$ is inversely proportional to the number of frequency
mesh points.\cite{Aulbur2000} The scaling problem is further compounded if
diagrammatic vertex corrections beyond the GW approximation are
included.\cite{Schindlmayr1998} The design of more efficient implementations
is therefore a pressing task.

In order to achieve a more favorable scaling, Rojas {\it et al.}\ recently
proposed an algorithm based on an alternative representation of the Green
function in real space and imaginary time.\cite{Rojas1995,Rieger1999} This
approach allows the calculation of the irreducible polarizability and the
self-energy as simple products rather than numerically expensive
convolutions. Furthermore, it directly exploits locality, i.e., the fact that
the Green function decays to zero as its spatial or temporal arguments move
apart. As a result, with a fixed cutoff energy this so-called space-time
method only scales like $N^2/\gamma^3 b$, where $\gamma$ denotes the
exponential decay constant in real space. Logarithmic terms are neglected.
Under the assumption that the frequency mesh maps one-to-one onto the relevant
region of the imaginary time axis, $b$ can there be identified with the
inverse period $\frac{1}{2} k_B T$ in a finite-temperature formalism and with
the exponential decay constant otherwise. The locality of the Green function
is a manifestation of the more general ``nearsightedness''
principle\cite{Kohn1996} and due to destructive quantum interference in
many-electron systems.

While the behavior of the Green function in reciprocal space and the
frequency domain has been analyzed in great detail,\cite{Farid1999} there has
so far been no comprehensive study of the decay properties in real space and,
for zero temperature, on the imaginary time axis. In fact, even qualitative
features are not always recognized correctly: Ref.\ \onlinecite{Aulbur2000},
for instance, describes the asymptotic tail of the Green function as being
proportional to $|{\bf r} - {\bf r}'|^{-2}$. While this assertion is correct
for the homogeneous electron gas at zero temperature, I argue here that real
materials in general exhibit exponential rather than algebraic decay. Attempts
to model the self-energy using results obtained for the homogeneous electron
gas,\cite{Sham1966,Wang1983,Sanchez-Friera2000} which are motivated by
seemingly universal features in the short-range part of the
nonlocality\cite{Godby1988} and the success of the local-density
approximation\cite{Kohn1965} in density-functional theory,\cite{Hohenberg1964}
should appreciate such fundamental differences.

The first objective of this paper is to systematically study the decay
properties of the one-particle Green function in solids, separately for metals
and insulators, and to obtain analytic expressions for the asymptotic behavior
in real space and imaginary time. Besides intrinsic interest, these can be
directly exploited in electronic-structure calculations. In the space-time
method, for instance, an analytic treatment of the asymptotic tail on the
imaginary time axis, in combination with a customized Gauss-Legendre grid,
reduces the number of mesh points where the Green function must be evaluated
numerically by one order of magnitude.\cite{Steinbeck2000} I show here that
the exponential fit proposed in Ref.\ \onlinecite{Steinbeck2000} is only valid
for systems with a finite band gap, however, while metals at zero temperature
instead require an algebraic fitting function. The second objective is to
relate the decay constants $\gamma$ and $b$ to known system parameters. This
is then used for a more detailed scaling comparison between the space-time
approach and conventional implementations.

Due to the central role of locality in $O(N)$ methods\cite{Goedecker1999}
within density-functional theory, the spatial decay rate of the one-particle
density matrix, which is a special element of the Green function, has recently
received some attention.\cite{Goedecker1998,Ismail-Beigi1999} Physically, the
nonlocality of the density matrix stems entirely from the delocalized quantum
character of the wave functions, while the Green function also describes the
actual transport of particles. The results obtained for the density matrix can
therefore not be generalized in a straightforward way, although they are
recovered here if the imaginary-time argument approaches zero from
below. Furthermore, the one-particle density matrix contains no information
about the behavior of the Green function on the imaginary time axis.

This paper is organized as follows. Sec.\ \ref{Sec:greenfunction} introduces
the Green function in real space and imaginary time. In Sec.\
\ref{Sec:electrongas} an analytic solution for the homogeneous electron gas is
presented. I derive asymptotic formulas for real metals and insulators in
Secs.\ \ref{Sec:metals} and \ref{Sec:insulators}. The results are summarized
in Sec.\ \ref{Sec:summary} together with a discussion. Atomic units are used
throughout.

\section{The Green function}\label{Sec:greenfunction}

The imaginary-time or thermal Green function is defined as\cite{Fetter1971}
\begin{eqnarray}
{\cal G}({\bf r},{\bf r}';\tau)
&=& \Theta(-\tau) \langle e^{(\hat{H}-\mu\hat{N})\tau} \hat{\psi}^\dagger({\bf
  r}') e^{-(\hat{H}-\mu\hat{N})\tau} \hat{\psi}({\bf r}) \rangle \\
&&- \Theta(\tau) \langle e^{-(\hat{H}-\mu\hat{N})\tau} \hat{\psi}({\bf r})
e^{(\hat{H}-\mu\hat{N})\tau} \hat{\psi}^\dagger({\bf r}') \rangle, \nonumber
\end{eqnarray}
where $\hat{\psi}^\dagger({\bf r})$ and $\hat{\psi}({\bf r})$ denote the
creation and annihilation operator for an electron at ${\bf r}$. The
Hamiltonian $\hat{H}$ is modified by the chemical potential $\mu$ and the
electron number operator $\hat{N}$, the angular brackets indicate the
thermodynamic average, and $\Theta(\tau)$ denotes Heaviside's step
function. Spin degrees of freedom are suppressed to simplify the
notation. At finite temperature the real variable $\tau$ is restricted to
$-\beta < \tau < 0$ with $\beta = 1 / k_B T$ for the hole part and $0 < \tau <
\beta$ for the electron part, and the Green function is periodically
repeated. Due to the additional antiperiodicity ${\cal G}(\tau) = -{\cal
  G}(\tau+\beta)$ it suffices to examine the hole part. At zero temperature
the period becomes infinite, however, and the properties of both parts must be
studied separately. Note that Refs.\ \onlinecite{Rieger1999,Steinbeck2000}
employ an alternative prefactor convention $G(i\tau) = i {\cal G}(-\tau)$ for
the imaginary-time Green function at zero temperature, which stems from
analytic continuation in the time domain rather than the frequency domain.

For real materials it is convenient to rewrite the hole part of the Green
function in the form
\begin{equation}\label{Eq:G_sum}
{\cal G}({\bf r},{\bf r}';\tau) = \sum_n  {\cal G}_n({\bf r},{\bf r}';\tau),
\end{equation}
\begin{equation}\label{Eq:Gn_sum}
{\cal G}_n({\bf r},{\bf r}';\tau) = \sum_{{\bf R},{\bf R}'} w_n({\bf
  r}-{\bf R}) F_n({\bf R}-{\bf R}';\tau) w^*_n({\bf r}'-{\bf R}'),
\end{equation}
\begin{equation}
w_n({\bf r}-{\bf R}) = \frac{1}{\Omega_B} \int\! e^{-i{\bf k}\cdot{\bf R}}
\psi_{n{\bf k}}({\bf r}) \,d^3k,
\end{equation}
\begin{equation}\label{Eq:Fn}
F_n({\bf R}-{\bf R}';\tau) = \frac{1}{\Omega_B} \int\! e^{i{\bf k}\cdot({\bf
    R}-{\bf R}')}
\frac{e^{-(\epsilon_{n{\bf k}}-\mu)\tau}}{1+e^{(\epsilon_{n{\bf
    k}}-\mu)\beta}} \,d^3k
\end{equation}
and analogous for the electron part. Here ${\bf R}$ labels the lattice
vectors, $w_n({\bf r}-{\bf R})$ indicates the Wannier orbitals corresponding
to a set of Bloch wave functions $\psi_{n{\bf k}}({\bf r})$, and the integrals
extend over the first Brillouin zone with volume $\Omega_B$. Wannier orbitals
are widely accepted to exhibit exponential localization. In addition to a
rigorous proof for isolated bands,\cite{Nenciu1983} there is compelling
numerical evidence to this effect for the general case of composite
bands.\cite{Marzari1997} Crucially, the localization persists even if defects
or surfaces break the periodicity of the solid.\cite{Nenciu1993} As the decay
length of Wannier orbitals is of the order of the interatomic spacing, only a
small number of terms contributes significantly to the sum (\ref{Eq:Gn_sum}).

\section{The homogeneous electron gas}\label{Sec:electrongas}

For the homogeneous electron gas, the hole part of the Green function is given
by
\begin{equation}\label{Eq:G_electrongas}
{\cal G}(r,\tau) = \frac{1}{8\pi^3} \int\! e^{i{\bf k}\cdot{\bf r}}
\frac{e^{-(\epsilon_k-\mu)\tau}}{1+e^{(\epsilon_k-\mu)\beta}} \,d^3k
\end{equation}
with $\epsilon_k = \frac{1}{2} k^2$. The range of integration can be extended
to infinity, because the spherical Fermi surface is fully contained inside the
Brillouin zone. The angular integrals are straightforward and lead to
\begin{equation}
{\cal G}(r,\tau) = -\frac{i}{4\pi^2r} \int_{-\infty}^\infty k e^{ikr}
\frac{e^{-(\epsilon_k-\mu)\tau}}{1+e^{(\epsilon_k-\mu)\beta}} \,dk.
\end{equation}
The remaining integral is most simply solved by closing the contour across the
upper complex half-plane. The integrand has relevant first-order poles at $\pm
k_l + i \gamma_l$ with
\begin{equation}
k_l = \sqrt{\mu + \sqrt{\mu^2 + \omega_l^2}},
\end{equation}
\begin{equation}
\gamma_l = \sqrt{-\mu + \sqrt{\mu^2 + \omega_l^2}},
\end{equation}
and the fermion Matsubara frequencies $\omega_l = (2l+1)\pi/\beta$. Evaluation
of the residues yields the final result
\begin{equation}
{\cal G}(r,\tau) = -\frac{1}{\pi \beta r} \sum_{l=0}^\infty e^{-\gamma_l r}
\cos(k_l r - \omega_l \tau),
\end{equation}
which explicitly shows the periodicity along the imaginary time axis. At
finite temperature the Green function is essentially a superposition of
exponentially damped oscillations, whose long-range behavior is dominated by
the $l=0$ contribution. For sufficiently small $T$ the chemical potential
approximately equals the Fermi energy $\epsilon_F = \frac{1}{2} k_F^2$. The
damping constants are given by $\gamma_l \approx \omega_l/k_F$ in this regime,
and the wave vectors of the corresponding oscillations become $k_l \approx
k_F$. The overall decay constant is hence proportional to the temperature and
given by $\gamma = \gamma_0 = \pi k_B T / k_F$.

As the limit $T \to 0$ is nontrivial, the Green function at zero temperature
is most simply calculated from Eq.\ (\ref{Eq:G_electrongas}) with a step
function instead of the Fermi distribution. The integral can be solved
analytically and yields
\begin{eqnarray}
{\cal G}(r,\tau)
&=& -\frac{\sin(k_F r)}{2 \pi^2 r \tau} + \frac{1}{2(2\pi|\tau|)^{3/2}}
\exp\left(\frac{k_F^2\tau}{2} - \frac{r^2}{2\tau} \right) \nonumber \\
&&\times \left[ \mbox{erfi}\left(\frac{k_F \tau + i r}{\sqrt{2|\tau|}} \right)
  + \mbox{erfi}\left(\frac{k_F \tau - i r}{\sqrt{2|\tau|}} \right) \right].
\end{eqnarray}
For the electron part $\mbox{erfc}(z) = 1 - \mbox{erf}(z)$ replaces
$\mbox{erfi}(z) = \mbox{erf}(iz)/i$. The limiting formulas are identical in
either case and follow from the asymptotic representation of the error
function $\mbox{erf}(z)$ in the complex plane. For $r \to \infty$ the Green
function becomes
\begin{equation}
{\cal G}(r,\tau) \sim -\frac{k_F \cos(k_F r)}{2 \pi^2 r^2} + \frac{1 - k_F^2
  \tau}{2 \pi^2 r^3} \sin(k_F r),
\end{equation}
which decays algebraically rather than exponentially. On the imaginary time
axis the Green function decays like
\begin{equation}
{\cal G}(r,\tau) \sim -\frac{\sin(k_F r)}{2 \pi^2 r \tau} - \frac{\cos(k_F
  r)}{2 \pi^2 k_F \tau^2}
\end{equation}
for large positive or negative $\tau$.

\section{Metals}\label{Sec:metals}

The explicit solution for the homogeneous electron gas illustrates more
general principles that also apply to real metals. Bands above or below the
Fermi level yield contributions to the Green function analogous to the case of
insulators, which is discussed below. They are short-ranged, so that for
$|{\bf r}-{\bf r}'| \to \infty$ the sum (\ref{Eq:G_sum}) is dominated by terms
arising from partially occupied bands that cross the Fermi level. Furthermore,
the strong localization of the Wannier orbitals implies that the spatial decay
properties of the relevant ${\cal G}_n$ are in turn determined by $F_n$. If
the homogeneous electron gas is used for guidance, then the decay length
$1/\gamma$ for sodium with an average density $r_s = 3.99$ (Ref.\
\onlinecite{Lide1995}) at room temperature $T = 298\,{\rm K}$ may be estimated
as 85.8$\,$\AA\@. In contrast, the decay length of the temperature-independent
Wannier orbitals is comparable to the interatomic spacing 3.72$\,$\AA\ in the
bcc cell and more than an order of magnitude smaller. The high-temperature
region where $F_n$ becomes sufficiently short-ranged such as not to dominate
the asymptotic behavior is irrelevant for solid-state physics, because it lies
above the melting point of the crystal lattice.

$F_n$ has the mathematical form of a three-dimensional Fourier transform and
is long-ranged because of rapid oscillations of the integrand in Eq.\
(\ref{Eq:Fn}). Quantitatively, at finite temperature the integrand falls from
near unity to zero in a narrow region around the Fermi surface. Its width is
given by $\delta k \propto k_B T / |\nabla\epsilon_{n{\bf k}}|$, where the
gradient is taken at the Fermi level. Directional effects due to the
anisotropic crystal structure are ignored here, and it is assumed that the
energy dispersion is analytic and smooth on the scale of $\delta k$. For
sufficiently small $T$ the latter condition is always fulfilled. According to
basic Fourier analysis,\cite{Bracewell1965} the extent of this rapid variation
is directly proportional to the exponential decay constant of the transform,
so that $\gamma \propto \delta k$ for $F_n$ and hence for ${\cal G}$.

At $T=0$ the integrand changes qualitatively, because the Fermi distribution
has a discontinuity at the Fermi surface. As the Fourier transform of a
discontinuous function decays algebraically rather than
exponentially,\cite{Bracewell1965} both the electron and the hole part of the
Green function now assume the asymptotic form ${\cal G}({\bf r},{\bf r}';\tau)
\sim |{\bf r}-{\bf r}'|^{-\eta}$ with $\eta > 0$. Their decay on the imaginary
time axis is likewise algebraic ${\cal G}({\bf r},{\bf r}';\tau) \sim
|\tau|^{-\kappa}$ with $\kappa > 0$, because the exponential
$\exp(-\epsilon_{n{\bf k}}\tau)$ in the integrand reaches the Fermi level and
thus exhausts the prefactor $\exp(\mu\tau)$.

\section{Insulators}\label{Sec:insulators}

Temperature effects are negligible for insulators if the thermal energy $k_B
T$ is much smaller than the band gap $\Delta$. As this condition is almost
always fulfilled below the melting point, it suffices to examine the case
$T=0$. If no band crosses the chemical potential, then all $F_n$ are spatially
short-ranged. For $\tau \to 0$ the integral (\ref{Eq:Fn}) has the
straightforward solution $F_n({\bf R}-{\bf R}';0) = \delta_{{\bf R},{\bf
    R}'}$. At finite imaginary times the only significant contribution comes
from the vicinity of the band edges. The energy dispersion can hence be
replaced by its harmonic approximation $|\epsilon_{n{\bf k}} - \mu| \approx
\frac{1}{2} \Delta + \frac{1}{2} \sum_{i,j} k_i (m^*)^{-1}_{ij} k_j$, where
$m^*$ denotes the effective mass tensor at the top of the valence band for the
hole part and at the bottom of the conduction band for the electron part. The
range of integration may now be extended to infinity, and $F_n$ is readily
seen to be of Gaussian type. The overexponential falloff implies that for
insulators the asymptotic properties of the Green function in real space are
determined by the Wannier orbitals, which are exponentially localized. In the
weak-binding case, where the band structure can be obtained by perturbation
from that of the homogeneous electron gas, their decay rate is given by
$\gamma \propto a \Delta$, where $a$ is the lattice
constant.\cite{Ismail-Beigi1999} Many common semiconductors have gaps
substantially smaller than their band width and are thus expected to fall into
the weak-binding regime, e.g., for silicon $\Delta = 1.15\,{\rm eV}$ (Ref.\
\onlinecite{Lide1995}) as opposed to $\frac{1}{2} (2\pi/a)^2 = 5.1\,{\rm
  eV}$. The proportionality factor is direction-dependent but about unity, so
that the decay length $1/\gamma$ may be estimated as 1.22$\,$\AA, which is of
the order of the interatomic spacing 2.35$\,$\AA\ in silicon, as expected, and
compatible with numerical calculations.\cite{Marzari1997} In the tight-binding
regime the Wannier orbitals also decay exponentially, but the dependence of
$\gamma$ on the gap is indeterminate and subject to details of the atomic
potential.\cite{Ismail-Beigi1999} This case may apply for large-gap
Mott-Hubbard insulators like NiO but is less relevant in practice, because the
GW approximation anyway shows serious deficiencies for such strongly
correlated systems.\cite{Aryasetiawan1996} The falloff on the imaginary time
axis is exponential with a decay constant $b = \frac{1}{2} \Delta$ both for
the electron and the hole part, because the finite difference between the
chemical potential and the band edges implies an incomplete cancellation of
the prefactor $\exp(\mu\tau)$.

\section{Summary}\label{Sec:summary}

In this paper I have investigated the asymptotic properties of the
one-particle Green function in real space and imaginary time. An analytic
solution for the homogeneous electron gas shows the Green function as a
superposition of damped harmonic oscillations with a wave vector approaching
$k_F$ as $T \to 0$. The damping ${\cal G} \sim \exp(-\gamma |{\bf r}-{\bf
  r}'|)$ is exponential at finite temperature but changes qualitatively to
algebraic falloff ${\cal G} \sim |{\bf r}-{\bf  r}'|^{-\eta}$ at $T=0$. While
the thermal Green function is by definition periodic on the imaginary time
axis with the period $2\beta$, it decays like ${\cal G} \sim |\tau|^{-\kappa}$
when the periodicity is lifted at $T=0$. I have shown that these results also
apply to real metals. In contrast, for insulators at zero temperature the
Green function decays exponentially like ${\cal G} \sim \exp(-\gamma |{\bf
  r}-{\bf r}'|)$ in real space and ${\cal G} \sim \exp(-b |\tau|)$ in
imaginary time.

The limiting formulas can be exploited in electronic-structure calculations
based on a space-time representation and allow an analytic treatment of the
long-range tails. The asymptotic behavior of derived propagators
within a perturbation scheme follows directly from the Green function. In
particular, the self-energy in the GW approximation is the product of
the Green function ${\cal G}$ and the screened Coulomb interaction $W$. In
insulators screening is incomplete with $W \sim 1/\epsilon |{\bf r}-{\bf
  r}'|$ at large distances, where $\epsilon$ is the dielectric constant, but
the self-energy still decays exponentially in real space because the Green
function does. On the other hand, for the homogeneous electron gas at zero
temperature\cite{Fetter1971} $W \sim \cos(2 k_F |{\bf r}-{\bf r}'|) /
|{\bf r}-{\bf r}'|^3$, so that the self-energy only decays algebraically. This
distinct asymptotic behavior should be appreciated if the electron-gas
self-energy is applied to real materials. The problem is naturally solved by
the {\it ad hoc}\/ introduction of a gap in the otherwise metallic
spectrum.\cite{Wang1983}

Finally, the generic dependence $\gamma \propto k_B T /
|\nabla\epsilon_{n{\bf k}}|$ and $b = \frac{1}{2} k_B T$ for metals as well as
$\gamma \propto a \Delta$ and $b = \frac{1}{2} \Delta$ for insulators can be
used to assess the scaling of different algorithms for self-energy
calculations in the GW approximation. If all other parameters remain fixed,
then conventional reciprocal-space implementations scale like $N^4/T^2$ for
metals at finite temperature and $N^4/\Delta^2$ for insulators at zero
temperature, whereas the space-time method scales like $N^2/T^4$ and
$N^2/\Delta^4$, respectively. This difference in efficiency gain as the
temperature or the band gap increase is an important factor to be taken into
account for benchmarking purposes.

\end{document}